\documentclass[10pt,a4paper]{article}
\usepackage[dvips]{color}
\usepackage{epsfig}
\usepackage{epstopdf}
\usepackage{amsmath}
\usepackage{graphicx}
\usepackage{authblk}
\usepackage{amssymb,amsmath}
\usepackage{amsmath}
\usepackage{tabularx}

\begin{document}
\title{\bf Some interacting dark energy models}
\author{{M. Khurshudyan$^{a,b,c}$\thanks{Email:khurshudyan@yandex.ru, khurshudyan@tusur.ru}~~and As. Khurshudyan$^{c}$\thanks{Email:khurshudyan@mechins.sci.am}}\\
$^{a}${\small {\em International Laboratory for Theoretical Cosmology, Tomsk State University of Control Systems and Radioelectronics (TUSUR), 634050 Tomsk, Russia}}\\ 
$^{b}${\small {\em Research Division,Tomsk State Pedagogical University, 634061 Tomsk, Russia}}\\
$^{c}${\small {\em Institute of Physics, University of Zielona Gora, Prof. Z. Szafrana 4a, 65-516 Zielona Gora,
Poland}}\\
$^{d}${\small {\em Institute of Mechanics, National academy of sciences (NAS) of Armenia, 24/2 Baghramyan ave., 0019 Yerevan, Armenia}}\\
}\maketitle

\begin{abstract}
In this paper we study new cosmological models involving new forms of non-gravitational interaction between cold dark matter and dark energy. The main purpose is to demonstrate the applicability of the forms of interaction to the problem in modern cosmology known as accelerated expansion of the large scale universe. It is well known, that a non-gravitational interaction can improve theoretical prediction, but a fundamental theory allowing to derive the form of the interaction is missing. Therefore, an active sought to find a correct form of the interaction from various phenomenological assumptions is still one of the active branches of research in modern cosmology. Besides cosmographic analysis, we perform $Om$ analysis.
\end{abstract}

\section{Introduction}\label{sec:INT}

One of the central ideas in modern cosmology is the non-gravitational interaction between dark energy and dark matter when attempting to explain the accelerated expansion of the large scale universe. On the other hand, according to general relativity there is no any restriction on the existence of interaction between other energy sources providing the background dynamics of our universe. Nevertheless, it is not clear, where non-gravitational interactions between two energy sources operating on different scales in our universe can arise from. We can assume for a while, that the origin of non-gravitational interaction is related to emergence of the spacetime dynamics. However, this is not of much help, since this hypothesis is not more fundamental compared with other phenomenological assumptions within modern cosmology~\cite{INT1}~-~\cite{INT13}~(and references therein).

The purpose of this paper is to develop new cosmological models, where new phenomenological forms of non-gravitational interactions are involved. Since in this paper we are interested in the problem of accelerated expansion of the large scale universe, we follow the well known approximation of the energy content of the recent universe~(for details we refer the reader to~\cite{MR} and references therein). Namely, we consider cold dark matter and barotropic dark fluid with negative constant equation of state parameter to represent the effective fluid with
\begin{equation}
P_{eff} = P_{de} = \omega_{de} \rho_{de},
\end{equation}
and
\begin{equation}
\rho_{eff} = \rho_{de} + \rho_{dm}.
\end{equation}
The need to have dark energy in order to provide a correct background dynamics consistent with the observational data is a well known fact and is discussed from the point when the accelerated expansion of the large scale universe was announced~\cite{MR}.

There are various models of dark energy and the variety is directly associated with the fact, that the tension between different datasets does not allow to choose one of them as the best candidate~\cite{DE}~(and references therein). However, the simplest model is cosmological constant with equation of state parameter $-1$. With this model we have additional problems, which can be solved with dynamical dark energy models, such as ghost dark energy and generalized holographic dark energy with Nojiri-Odintsov cut-off to mention a few~\cite{DE1}~-~\cite{DE9}. Dark fluid interpretation of dark energy is another approach. There is a systematic update in this direction and one of them is the varying polytropic fluid presented in Ref.~\cite{DE5}~(see the references therein about other models of dark fluids). The accelerated expansion of the large scale universe can be explained by modifying the general relativity~\cite{MG1}~-~\cite{MG12}. On the other hand, it can be done by particle creation, which generates negative pressure. Different modifications of general relativity and different models for parametrization of the particle creation rate has been considered in recent literature. We refer our readers to some of them~\cite{PC1}~-~\cite{PC5}. In general, a model can be constrained using the background tests and the growth test. In this work we will concentrate our attention only on the background tests involving the following four datasets:
\begin{enumerate}
\item The differential age of old galaxies, given by $H(z)$.
\item The peak position of baryonic acoustic oscillations (BAO).
\item The SN Ia data.
\item Strong Gravitation Lensing data.
\end{enumerate} 
In the case of the Observed Hubble Data, one defines chi-square given by
\begin{equation}
\chi^{2}_{OHD} = \sum \frac{\left ( H(\textbf{P},z) - H_{obs}(z)\right )^{2}}{\sigma_{OHD}^{2}},
\end{equation}
where $H_{obs}(z)$ is the observed Hubble parameter at redshift $z$ and $\sigma_{OHD}$ is the error associated with that particular observation, while $ H(\textbf{P},z)$ is the Hubble parameter obtained from the model and $\textbf{P}$ is the set of the parameters to be determined/constrained from the dataset. On the other hand, $7$ measurements have been jointly used determining the BAO~(Baryon Acoustic Oscillation) peak parameter to constrain the models by
\begin{equation}
\chi^{2}_{BAO} = \sum \frac{\left ( A(\textbf{P},z) - A_{obs}(z)\right )^{2}}{\sigma_{BAO}^{2}},
\end{equation}
where the theoretical value for the $\textbf{P}$ set of the parameters $A(\textbf{P},z)$ is determined as
\begin{equation}
A(\textbf{P},z_{1}) = \frac{\sqrt{\Omega_{m}}}{E(z_{1})^{1/3}} \left( \frac{ \int_{0}^{z_{1}} \frac{dz}{E(z)} }{z_{1}}\right)^{2/3},
\end{equation}
with $E(z) = H(z)/H_{0}$ and $H_{0}$ is the value of the Hubble parameter at $z=0$. For the Supernovae Data, $\chi^{2}_{\mu}$ is defined as
\begin{equation}
\chi^{2}_{\mu} = A - \frac{B^{2}}{C}, 
\end{equation}
where
\begin{equation}
A = \sum {\frac{(\mu(\textbf{P},z) - \mu_{obs})^{2}}{\sigma_{\mu}^{2}}}, 
\end{equation}
\begin{equation}
B = \sum {\frac{\mu(\textbf{P},z) - \mu_{obs}}{\sigma_{\mu}^{2}}} ,
\end{equation}
and
\begin{equation}
C = \sum {\frac{1}{\sigma_{\mu}^{2}}}.
\end{equation}
In the last $3$ equations $\sigma_{\mu}$ is the uncertainty in the distance modulus~\cite{DATA1}.

If the analysis is carried out by including the Strong Gravitational Lensing data, we must follow to the receipt of Ref.~\cite{DATA2} and use the data identical to the data presented there. The receipt of Ref.~\cite{DATA2} allows to impose observational constraints on the parameters of the models, without considering the structure and physics of the lensing object. To obtain appropriate constraints, usually known as the best fit values of the parameters of the model, one need to minimize $\chi^{2}$ function
\begin{equation}
\chi^{2} = \chi^{2}_{OHD} + \chi^{2}_{BAO}+ \chi^{2}_{\mu} + \chi^{2}_{SGL},
\end{equation} 
when all the datasets are used simultaneously (for each combination of the datasets appropriate total $\chi^{2}$ should be considered). The set of the parameters to be constrained by $\chi^{2}$ analysis for each model of this paper will be presented in the next section during the discussion of the models.

The paper is organized as follows: in section~\ref{MPSA} we discuss the models and perform cosmographic analysis to demonstrate their viability. At the same time we apply the well known $\chi^{2}$ statistical analysis to constraint the models. The cosmographic analysis is performed taking into account the constraints obtained on the parameters of the models. The description of the models includes also a presentation of the forms of non-gravitational interactions. In section~\ref{sec:Dis} we organize discussion on obtained results and present possible extensions of the models considered in this work.

\section{Models and observational constraints}\label{MPSA}

For the models considered in this paper we assume, that the general relativity describes the background dynamics. Moreover, we consider a flat low redshift Universe with FRW metric and interacting dark components, for which the field equations read as
\begin{equation}\label{eq: Fridmman vlambda}
H^{2}=\frac{\dot{a}^{2}}{a^{2}}=\frac{\rho}{3},
\end{equation}
\begin{equation}\label{eq:Freidmann2}
\frac{\ddot{a}}{a}= - \frac{1}{6}(\rho + 3 P).
\end{equation}
If we additionally assume, that the effective fluid is ideal, we derive the following equations describing the dynamics of cold dark matter and dark energy:
\begin{equation}\label{eq:inteqm}
\dot{\rho}_{dm}+3H \rho_{dm} = Q,
\end{equation}
\begin{equation}\label{eq:inteqG}
\dot{\rho}_{de}+3H\rho_{de} (1 + \omega_{de} )= -Q.
\end{equation}
These equations describe non-gravitational interaction providing a transition from dark energy to dark matter, while $\omega_{de}$ is a negative constant~(equation of state parameter of dark energy). The cosmographical analysis of the models very often are performed using $Om$ analysis. It is well known, that $Om$ analysis is a geometrical tool to study dark energy models involving the following parameter~\cite{Om}:
\begin{equation}
Om = \frac{x^{2}-1}{(1+z)^{3} - 1},
\end{equation}  
where $x = H/H_{0}$. Note, that the $Om$ analysis has been generalized to the two point $Om$ analysis with
\begin{equation}
Om(z_{2},z_{1}) = \frac{x(z_{2})^{2} - x(z_{1}^{2})}{(1+z_{2})^{2} - (1+z_{1})^{2}}.
\end{equation}
Moreover, a slight modification of the two point $Om$ ($Omh^{2}$) is suggested in Ref.~\cite{Omh} and the estimated values of the two point $Omh^{2}$ for $z_{1} = 0$, $z_{2} = 0.57$ and $z_{3} = 2.34$:
$$Omh^{2}(z_{1};z_{2}) = 0.124 \pm 0.045,$$
$$Omh^{2}(z_{1};z_{3}) = 0.122 \pm 0.01,$$
\begin{equation}
Omh^{2}(z_{2};z_{3}) = 0.122 \pm 0.012,
\end{equation}
are used in this paper to obtain constraints on the parameters of the models.

Recall, that for the $\Lambda$CDM mode, $Omh^{2}=0.1426$. The constraints on $Omh^{2}$ can be used to constrain the parameters of the models as well. The $Om$ analysis is the simplest tool to study dark energy models, since it connects the Hubble parameter and the redshifts. For other tools, like statefinder hierarchy analysis, we need to calculate higher order derivatives of the scale factor, which makes calculations costly and complicated. To simplify our discussion we organize $2$ sections discussing the main results obtained from the $\chi^{2}$ technique.

\section{Model 1}\label{sec:M1}
For the first cosmological model considered in this paper, we assume that the form of the non-gravitational interaction between dark energy and cold dark matter reads as
\begin{equation}\label{eq:Q1}
Q = 3 H b \rho_{de} \log \left[ \frac{\rho_{de}}{\rho_{dm}}\right],
\end{equation}
where $b$ is a constant and should be determined from the observational data. On the other hand, $\rho_{de}$ and $\rho_{dm}$ are the densities of dark energy and dark matter. Suggested form of the non-gravitational interaction is constructed from the classical interaction term $Q = 3Hb \rho_{de}$ intensively considered in literature. According to Eq.~(\ref{eq:inteqm}) and Eq.~(\ref{eq:inteqG}) the form of the interaction Eq.~(\ref{eq:Q1}) indicates an energy transition from dark energy to dark matter. On the other hand, the transition from dark matter to dark energy can be modeled by
\begin{equation}\label{eq:Q2}
Q = 3 H b \rho_{de} \log \left[ \frac{\rho_{dm}}{\rho_{de}}\right].
\end{equation}

\subsection{Transition from dark energy to dark matter}\label{ss:11}

Consideration of the interaction Eq.~(\ref{eq:Q1}) provides a cosmological model~(including the parameters from the other assumptions) with the following $\{H_{0},\Omega^{(0)}_{dm},\omega, b \}$ parameters, where $\omega$ is the equation of state parameter of dark energy, while $\Omega^{(0)}_{dm} = \rho^{(0)}_{dm}/3H^{2}_{0}$ and satisfies the following constraint
\begin{equation}
\Omega^{(0)}_{dm} + \Omega^{(0)}_{de} = 1.
\end{equation}
Consideration on only energy transition from dark energy and dark matter impose on the parameter $b$ to be strictly positive. This prior constrain on the parameter $b$ is taken into account to constrain the other parameters of the model using the datasets and described statistical $\chi^{2}$ analysis presented above. On the other hand, to reduce the number of the parameters and the computational time, for  
$\Omega^{(0)}_{dm}$ we consider $0.26$, $0.27$, $0.28$ values. Taking into account the results from the existing in scientific literature analysis, for instance, a prior constraints on $\omega$ and $H_{0}$ are $[-1.2,-0.3]$ and $[67.5,73.5]$, respectively. With such priors we obtain the best fit of the model with observational data described above as follows: $\{70.57,0.26,-1.03, 0.0 \}$, $\{70.22,0.27,-0.99, 0.062 \}$ and $\{69.88,0.28,-0.97, 0.105 \}$ with $\chi^{2} = 784.2$, $\chi^{2} = 780.9$ and $\chi^{2} = 779.6$, respectively.

\begin{figure}[h!]
 \begin{center}$
 \begin{array}{cccc}
\includegraphics[width=80 mm]{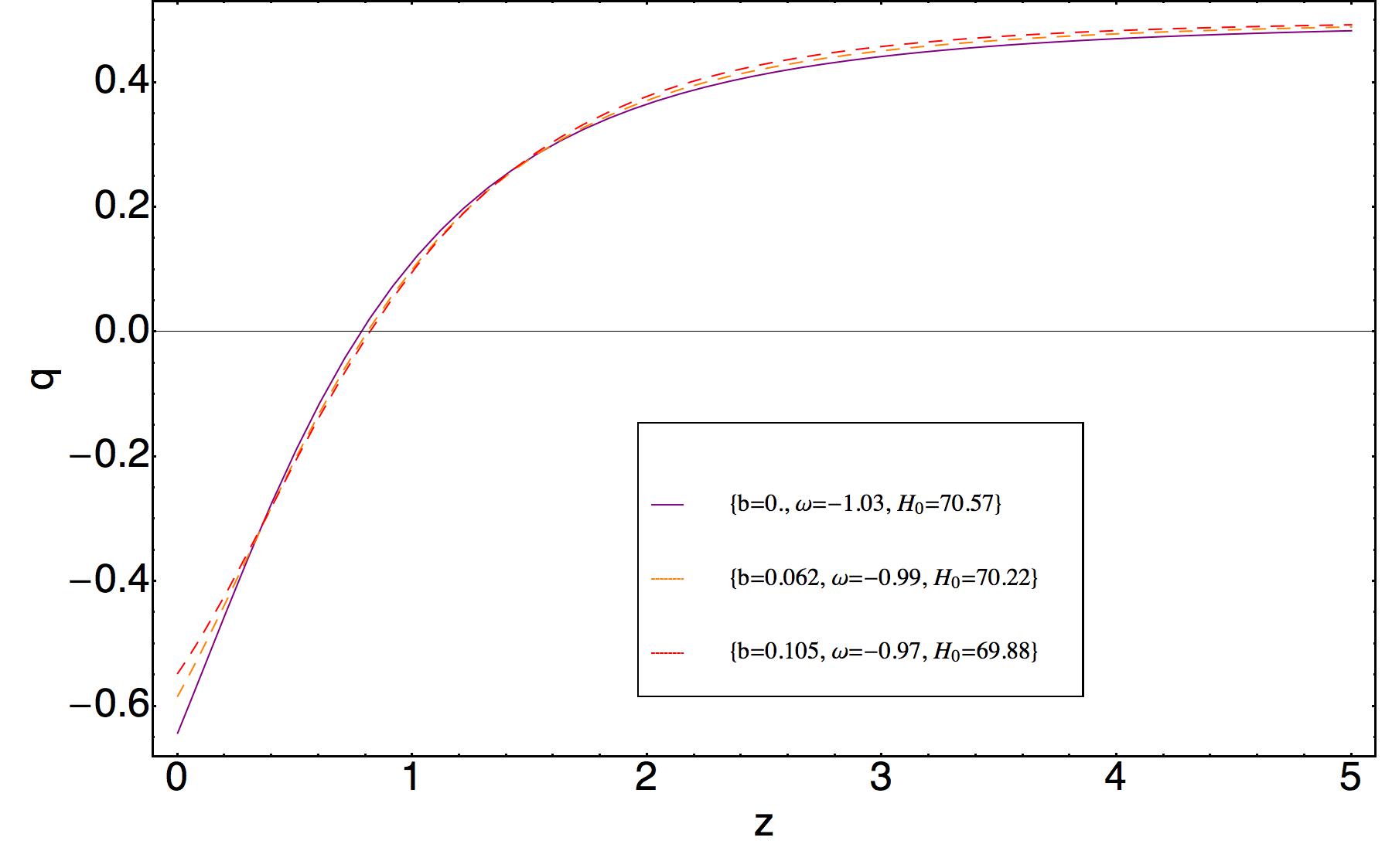}  &
\includegraphics[width=80 mm]{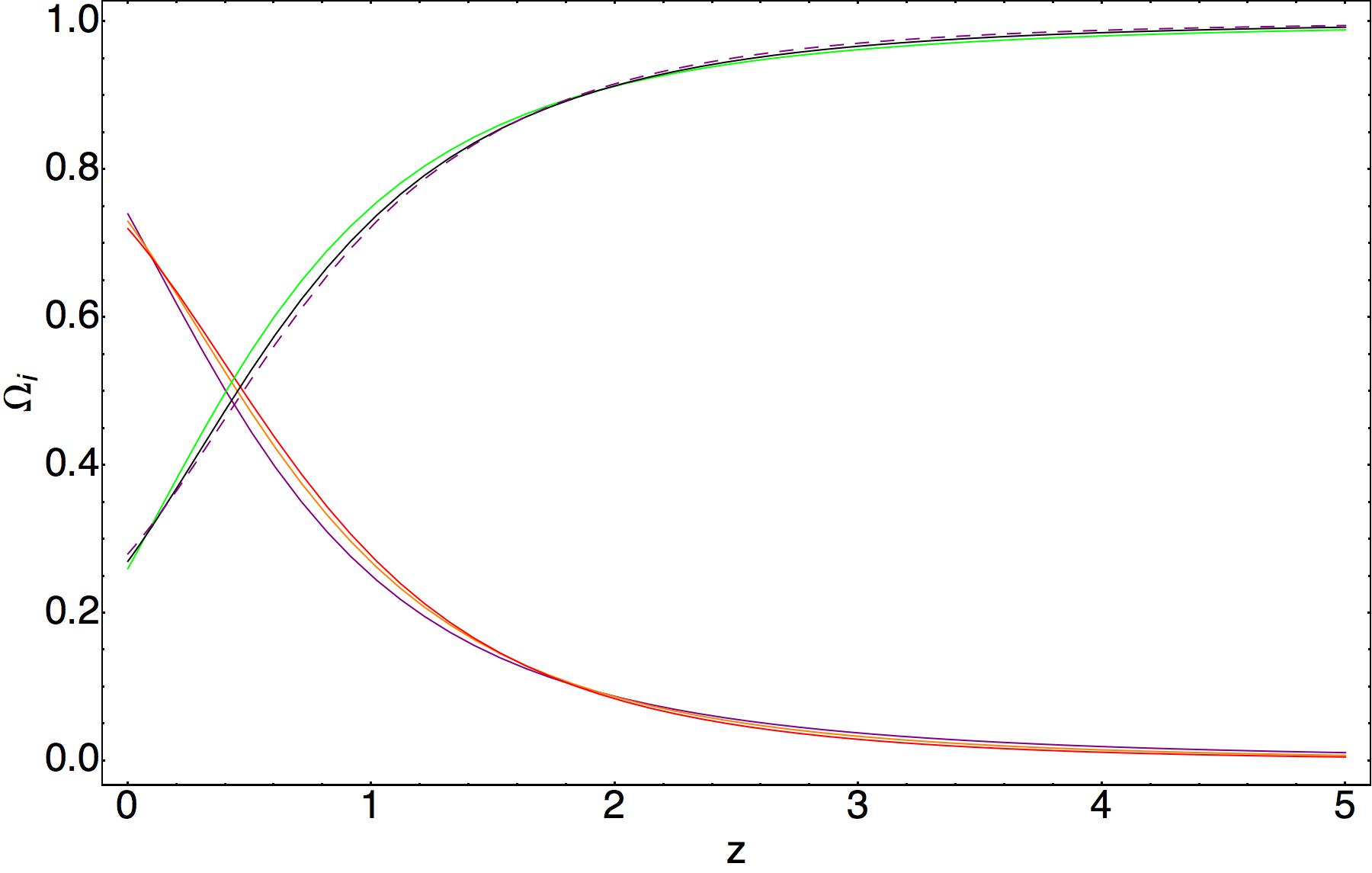}  \\
 \end{array}$
 \end{center}
\caption{Graphical behavior of the deceleration parameter $q$, $\Omega_{de}$~(solid lines) and $\Omega_{dm}$~(dashed lines) for the universe with two component fluid, when the non-gravitational interaction is given by Eq.~(\ref{eq:Q1})}
 \label{fig:Fig1}
\end{figure}

\begin{figure}[h!]
 \begin{center}$
 \begin{array}{cccc}
\includegraphics[width=80 mm]{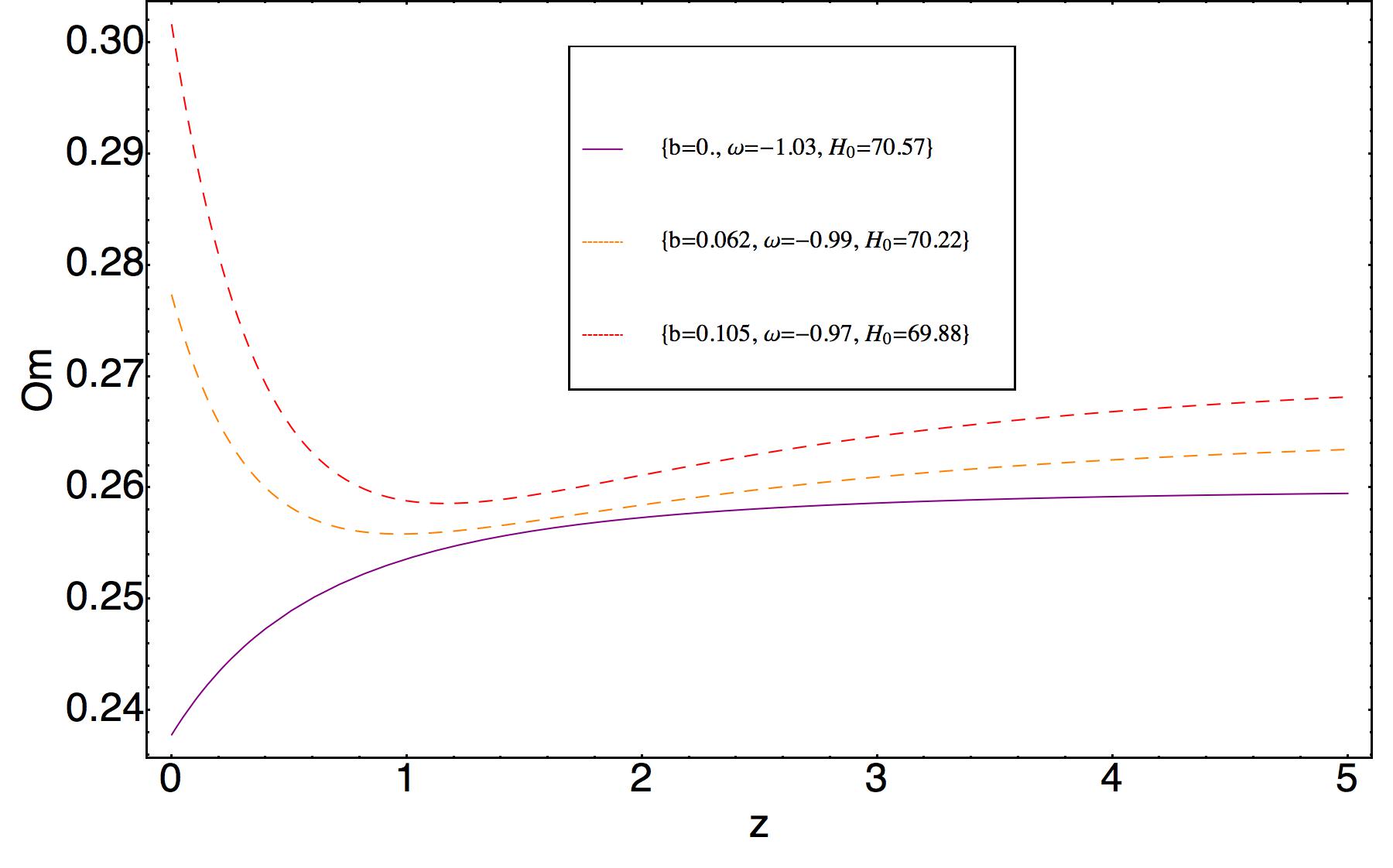}  &
\includegraphics[width=80 mm]{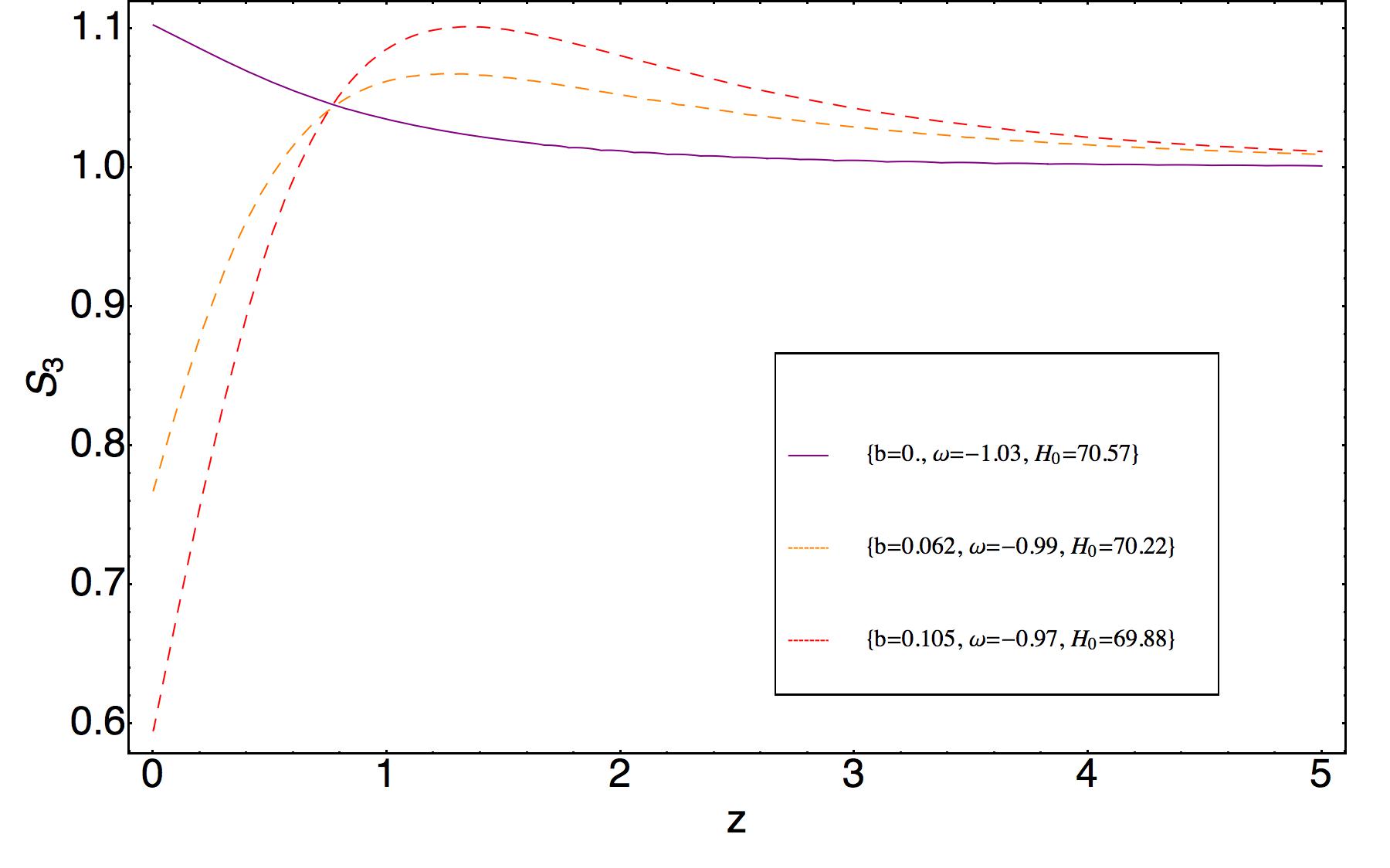}  \\
 \end{array}$
 \end{center}
\caption{Graphical behavior of $Om$ and $S_{3}$ parameters for the universe with two component fluid, when the non-gravitational interaction is given by Eq.~(\ref{eq:Q1})}
 \label{fig:Fig2}
\end{figure}

The graphical behavior of the deceleration parameter with the best fit values for the parameters is presented in Fig.~(\ref{fig:Fig1})~(the left plot). Transition between decelerated and accelerated expanding phases occurs. Moreover, the best fit results corresponding to $\Omega^{(0)}_{dm} = 0.27$ and $\Omega^{(0)}_{dm} = 0.28$ shows a slightly early transition between mentioned expanding phases, than the case with $\Omega^{(0)}_{dm} = 0.26$. On the other hand, since for the parameter space scanning we use the same priors and the same parameter space discretizing, we conclude that the model explains available/considered data~(best fit with $\chi^{2} = 779.6$) with $\{69.88,0.28,-0.97, 0.105\}$. At the same time, the right plot of Fig.~(\ref{fig:Fig1}) demonstrates, that the model is free from the cosmological coincidence problem. The graphical behaviors of $Om$ and $S_{3}$ parameters representing $Om$ and statefinder hierarchy analysis~(see~\cite{S3} for the definition) are presented in Fig.~(\ref{fig:Fig2}). We see clearly, that both parameters are very sensitive on the values of the model parameters. Moreover, they indicate clearly departures from the $\Lambda$CDM model. 

\subsection{Transition from dark matter to dark energy}

In this subsection we discuss the results corresponding to the best fit for the cosmological model, where the non-gravitational interaction between dark energy and dark matter is given by Eq.~(\ref{eq:Q2}). According to the assumption about existence of the interaction between dark energy and dark matter, energy transfer from dark matter to dark energy is indicated, unlike the interaction given by Eq.~(\ref{eq:Q1}). During the study of this case, the prior constraint on the parameter $b$ is extended. In particular, we allowed $b$ to be negative as well, which in this case also indicates transition from dark energy to dark matter corresponding to the case discussed in subsection~\ref{ss:11}. In this case, $-0.99$, $-1.0$, $-1.02$, $-1.1$ and $-1.2$ discrete priors on $\omega$ are imposed and the following constraints $b=\{-0.095,-0.084,-0.062,0.24,0.3\}$, $H_{0}=\{69.86,69.76,69.65,69.35,70.68\}$ and $\Omega^{(0)}_{dm} =\{0.28,0.28,0.28,0.28,0.28\}$ with $\chi^{2}=\{779.7,779.9,780.4,780.1,780.9\}$ are obtained.

\begin{figure}[h!]
 \begin{center}$
 \begin{array}{cccc}
\includegraphics[width=80 mm]{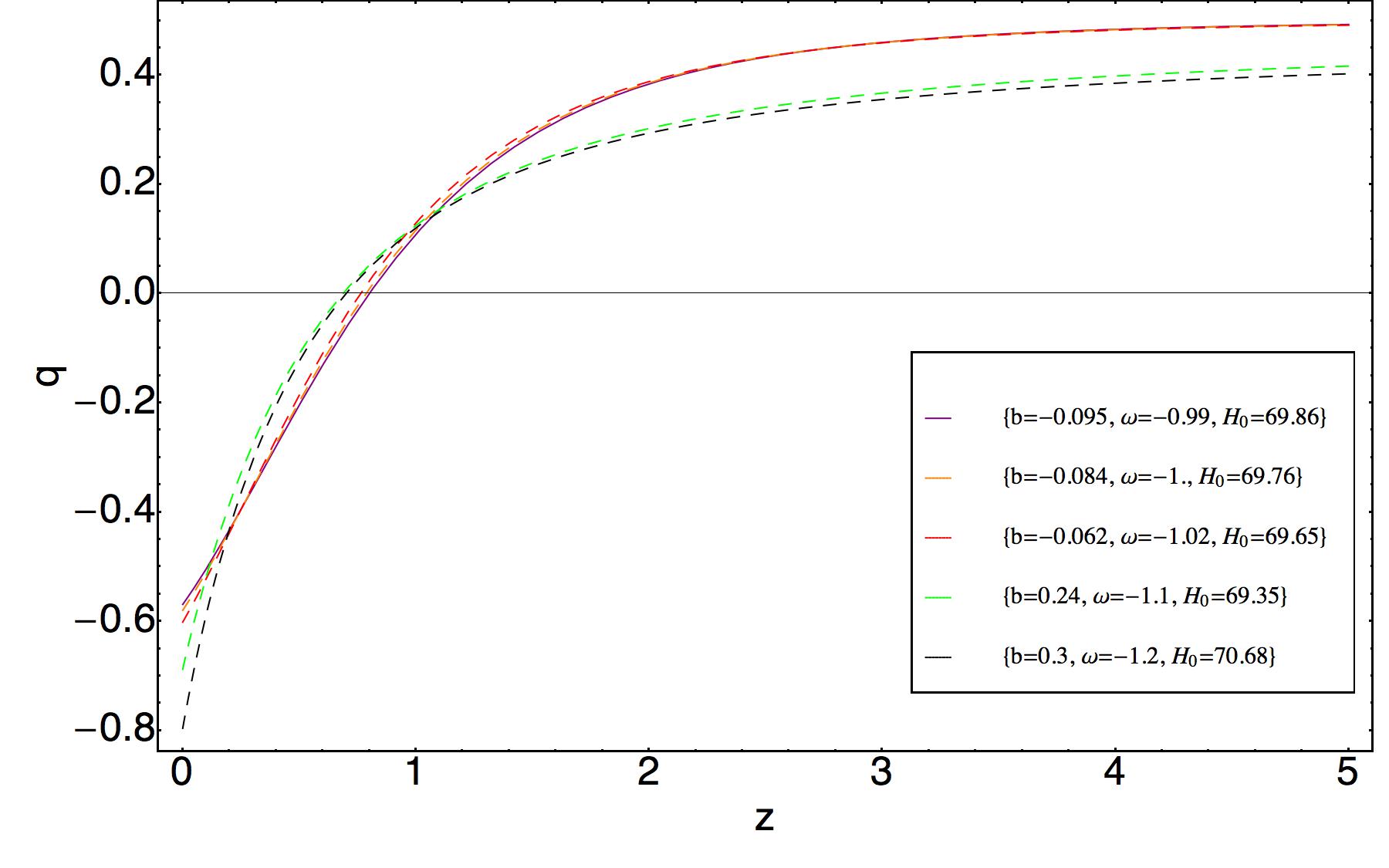}  &
\includegraphics[width=80 mm]{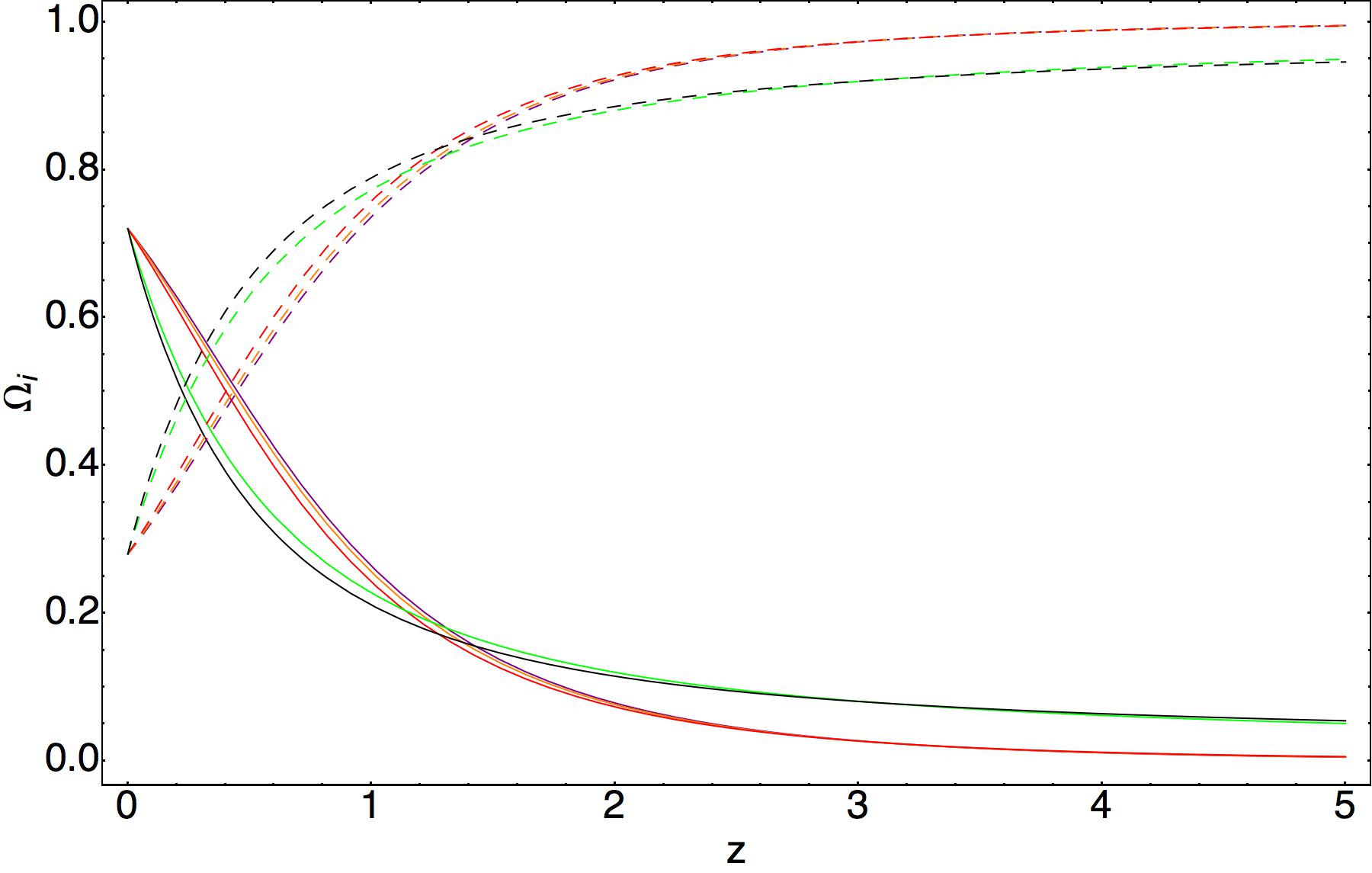}  \\
 \end{array}$
 \end{center}
\caption{Graphical behavior of the deceleration parameter $q$, $\Omega_{de}$~(solid lines) and $\Omega_{dm}$~(dashed lines) for the universe with two component fluid, when the non-gravitational interaction is given by Eq.~(\ref{eq:Q2})}
 \label{fig:Fig3}
\end{figure}

\begin{figure}[h!]
 \begin{center}$
 \begin{array}{cccc}
\includegraphics[width=80 mm]{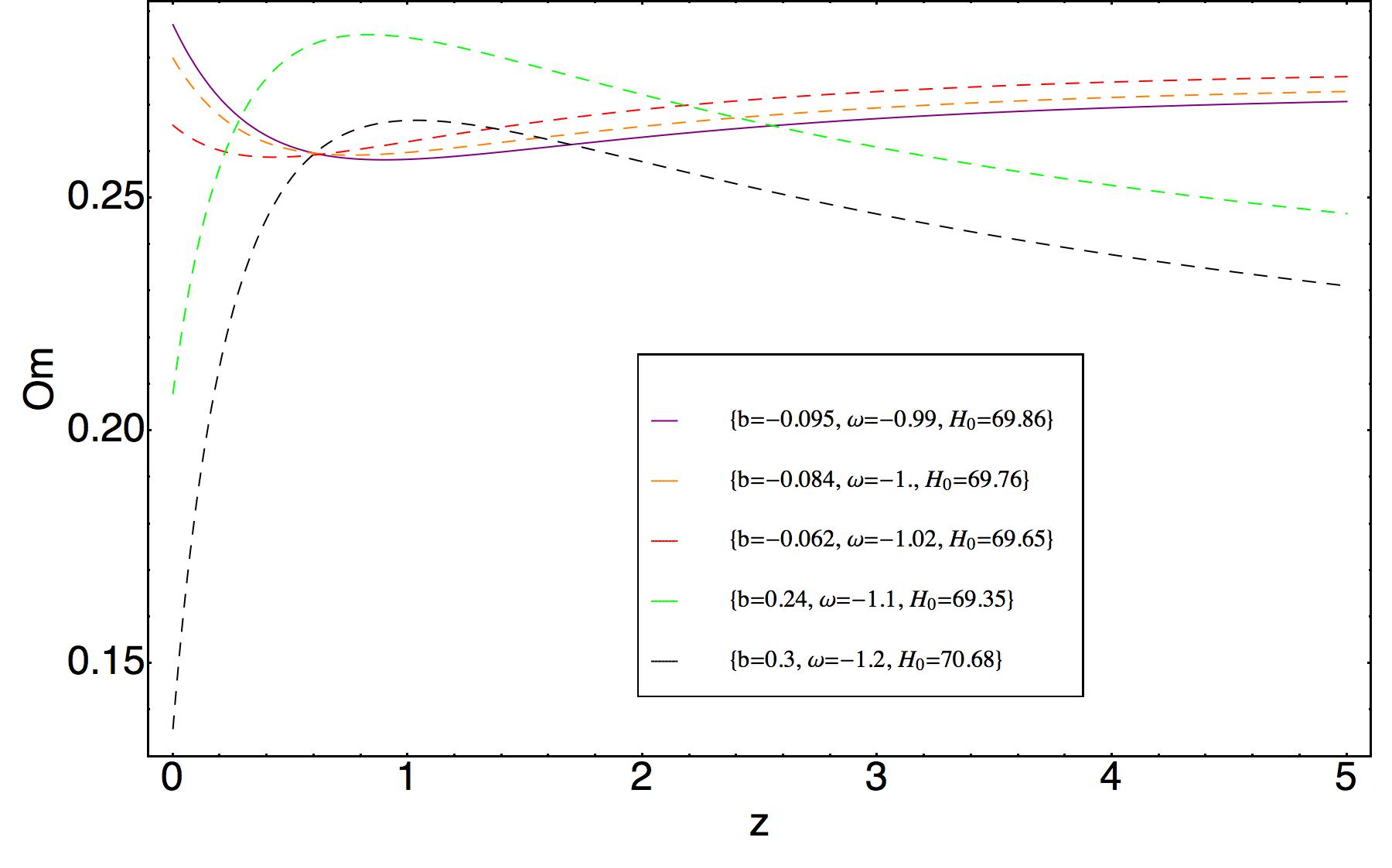}  &
\includegraphics[width=80 mm]{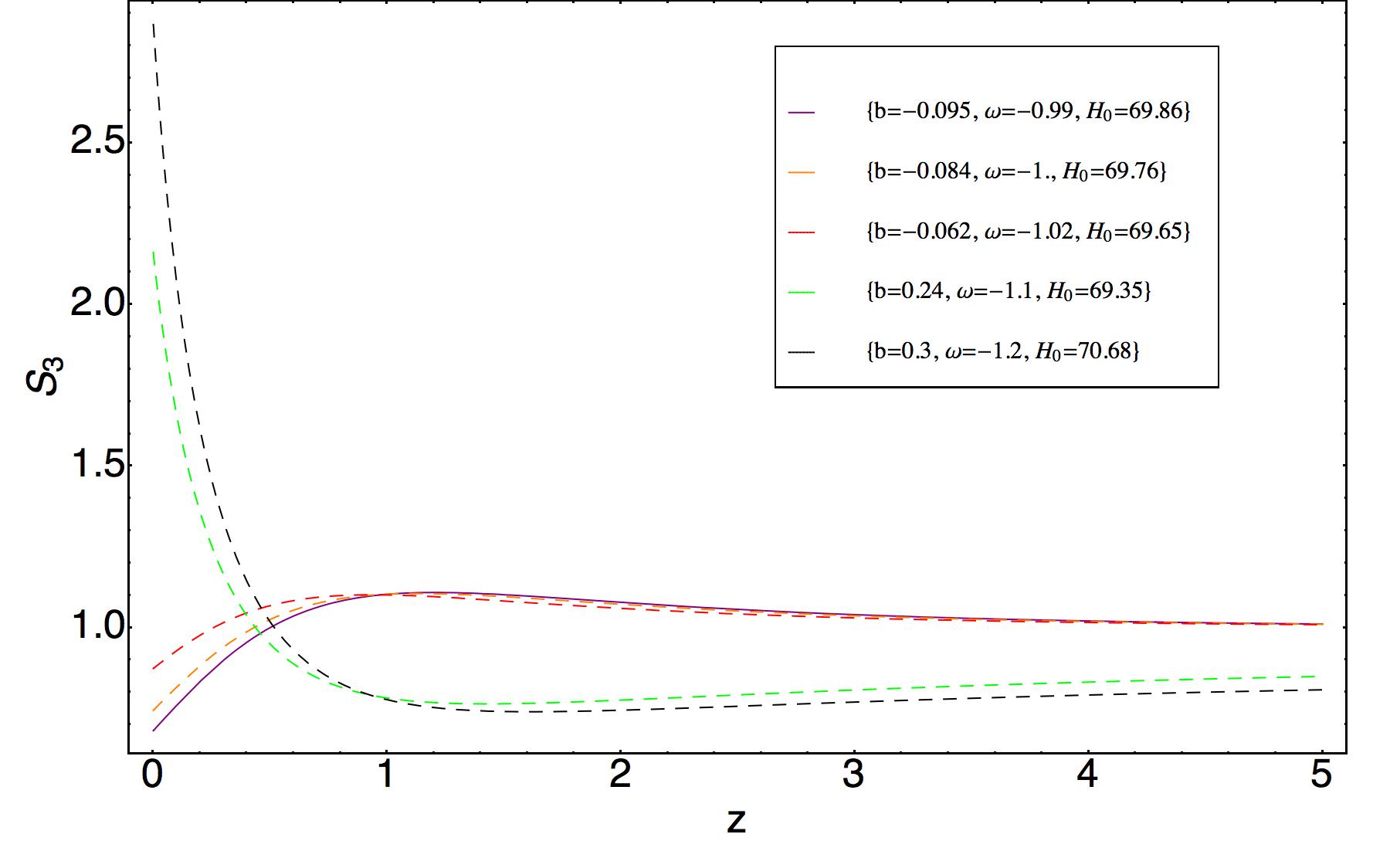}  \\
 \end{array}$
 \end{center}
\caption{Graphical behavior of $Om$ andf $S_{3}$ parameters for the universe with two component fluid, when the non-gravitational interaction is given by Eq.~(\ref{eq:Q2})}
 \label{fig:Fig4}
\end{figure}

From the obtained results we see, that when dark energy is quintessence with allowed lower value for the equation of state parameter supported by PLANCK 2015~\cite{PL}, then the considered data support only energy transition from dark energy to dark matter: due to the fact, that $b < 0$ in Eq.~(\ref{eq:Q2}). However, if with the parotropic fluid equation we will attempt to obtain a phantom dark energy universe, then it is possible only if transition for the energy will be from dark matter to dark energy, i.e. $b>0$. Graphical behaviors of the deceleration parameter $q$, $\Omega_{de}$, $\Omega_{dm}$ are presented in Fig.~(\ref{fig:Fig3}). The graphical behavior of $Om$ and $S_{3}$ parameters are presented in Fig.~(\ref{fig:Fig4}).

\section{Model 2}\label{sec:M2}

\begin{figure}[h!]
 \begin{center}$
 \begin{array}{cccc}
\includegraphics[width=80 mm]{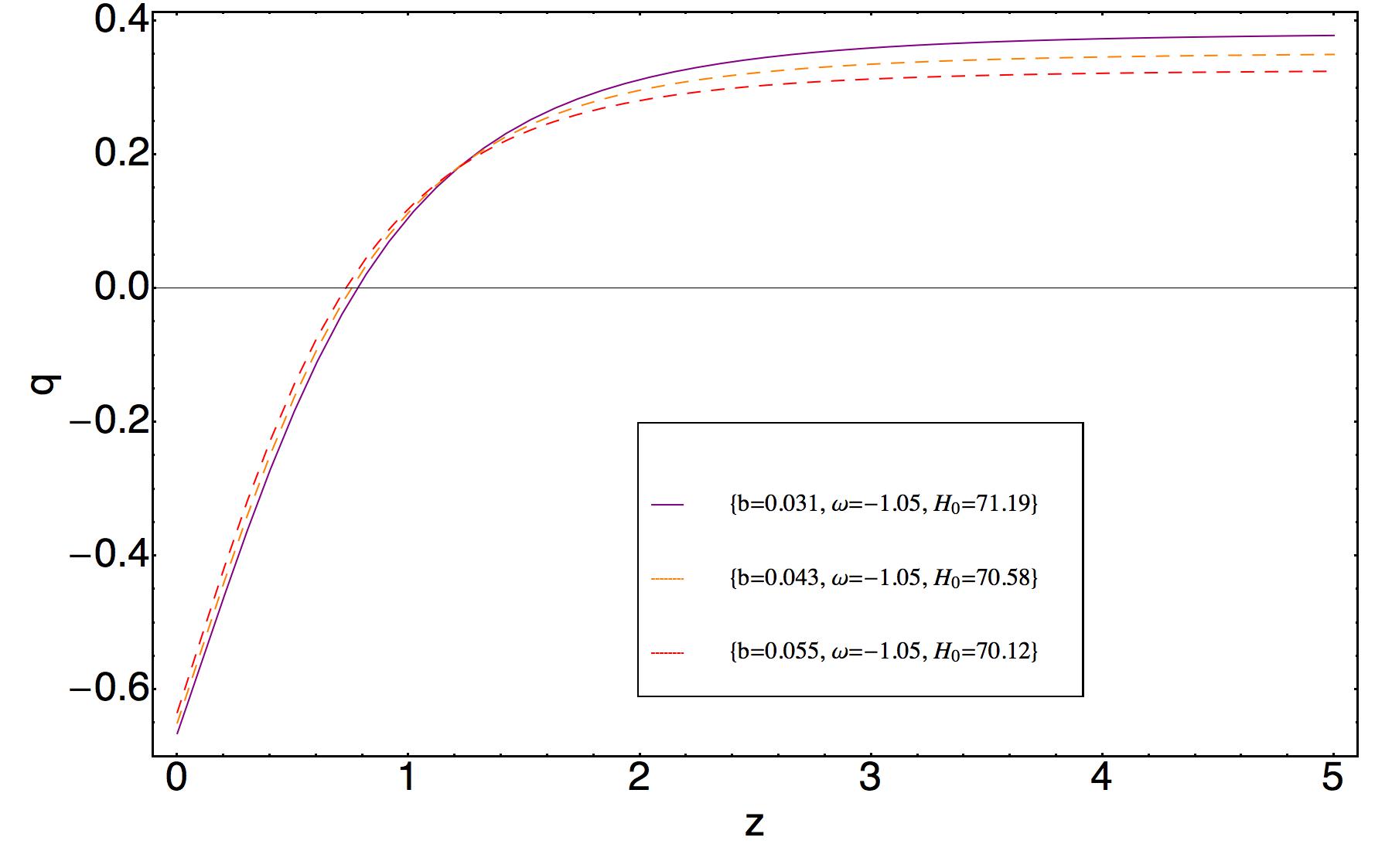}  &
\includegraphics[width=80 mm]{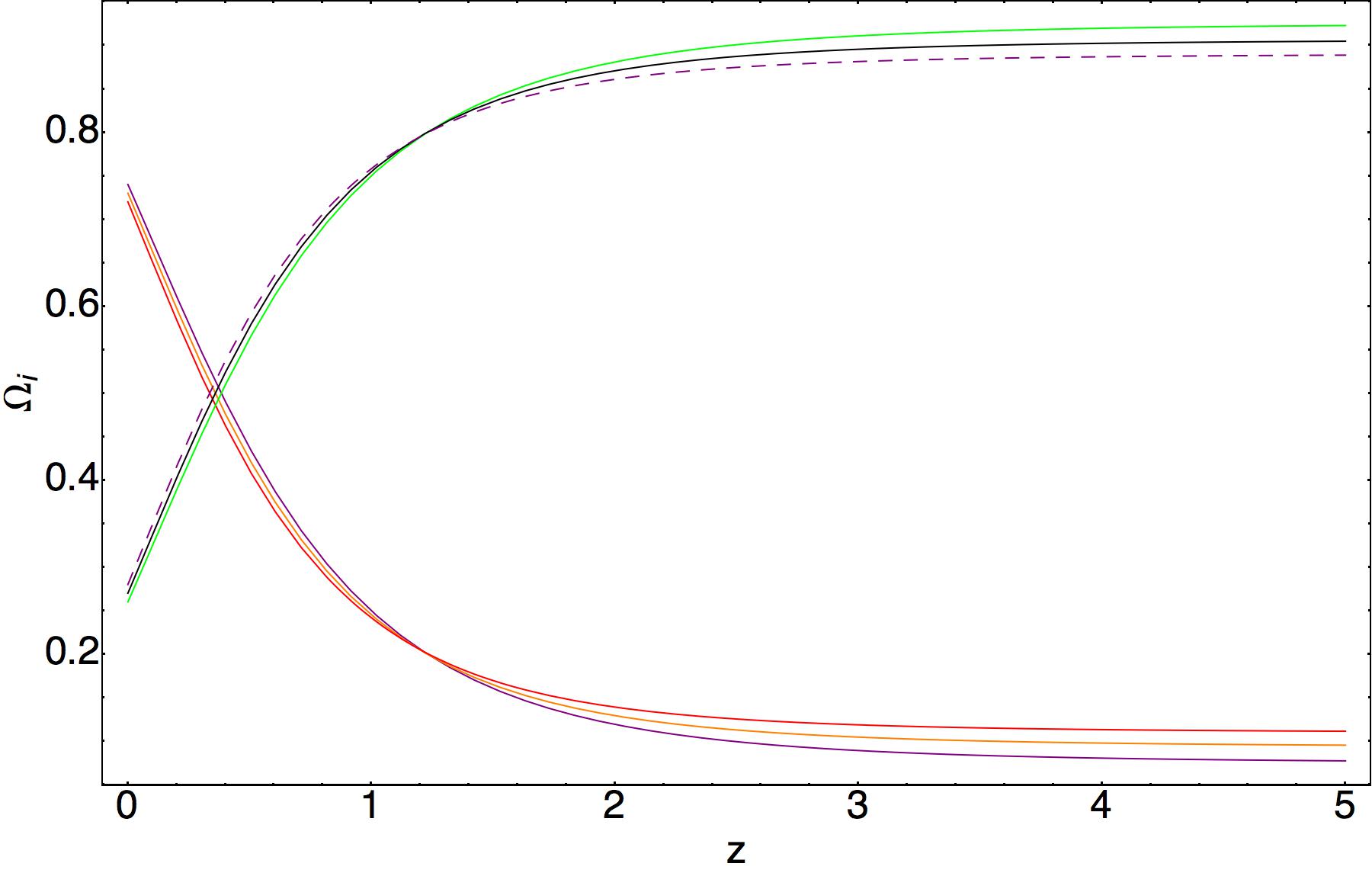}  \\
\includegraphics[width=80 mm]{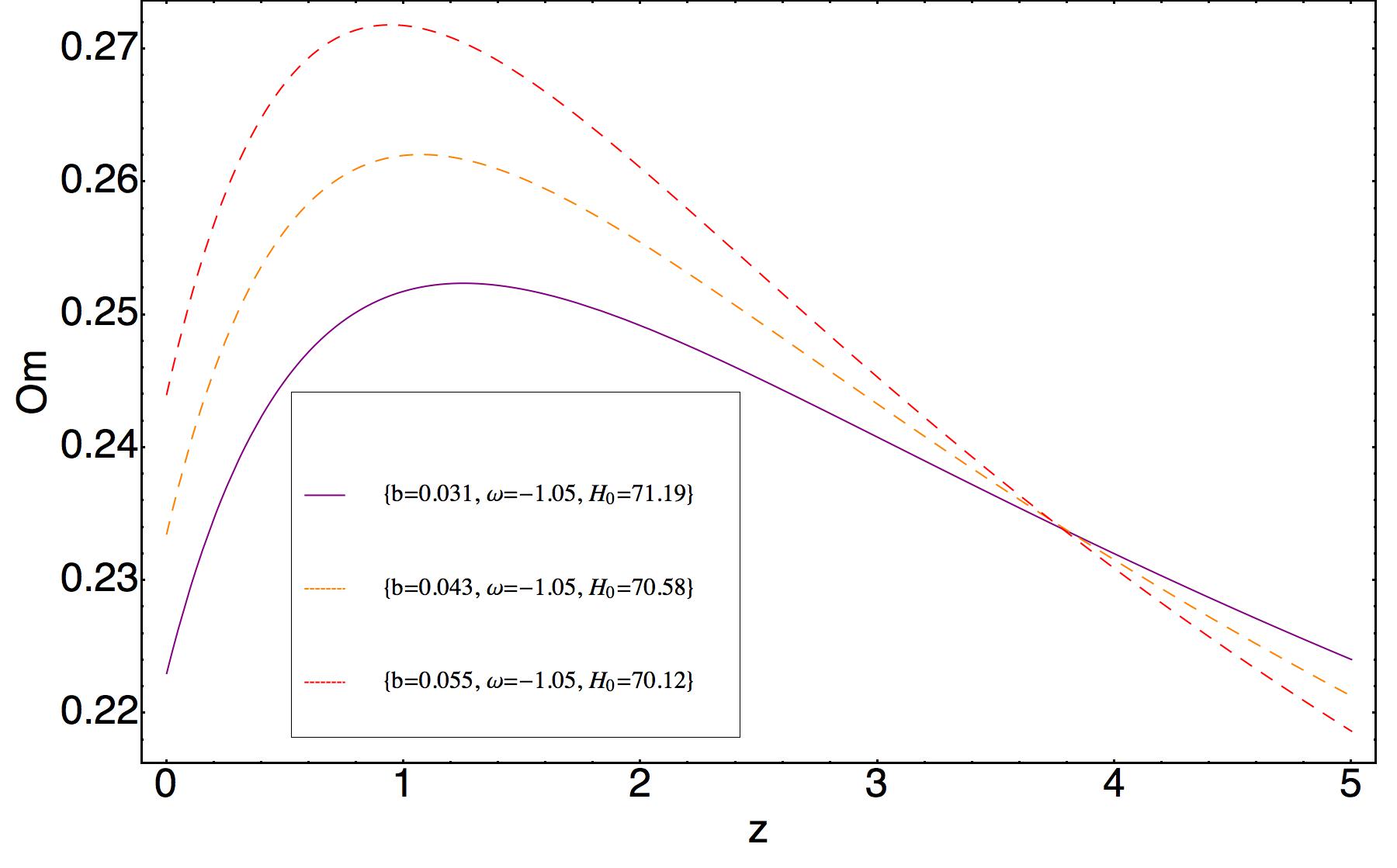}  &
\includegraphics[width=80 mm]{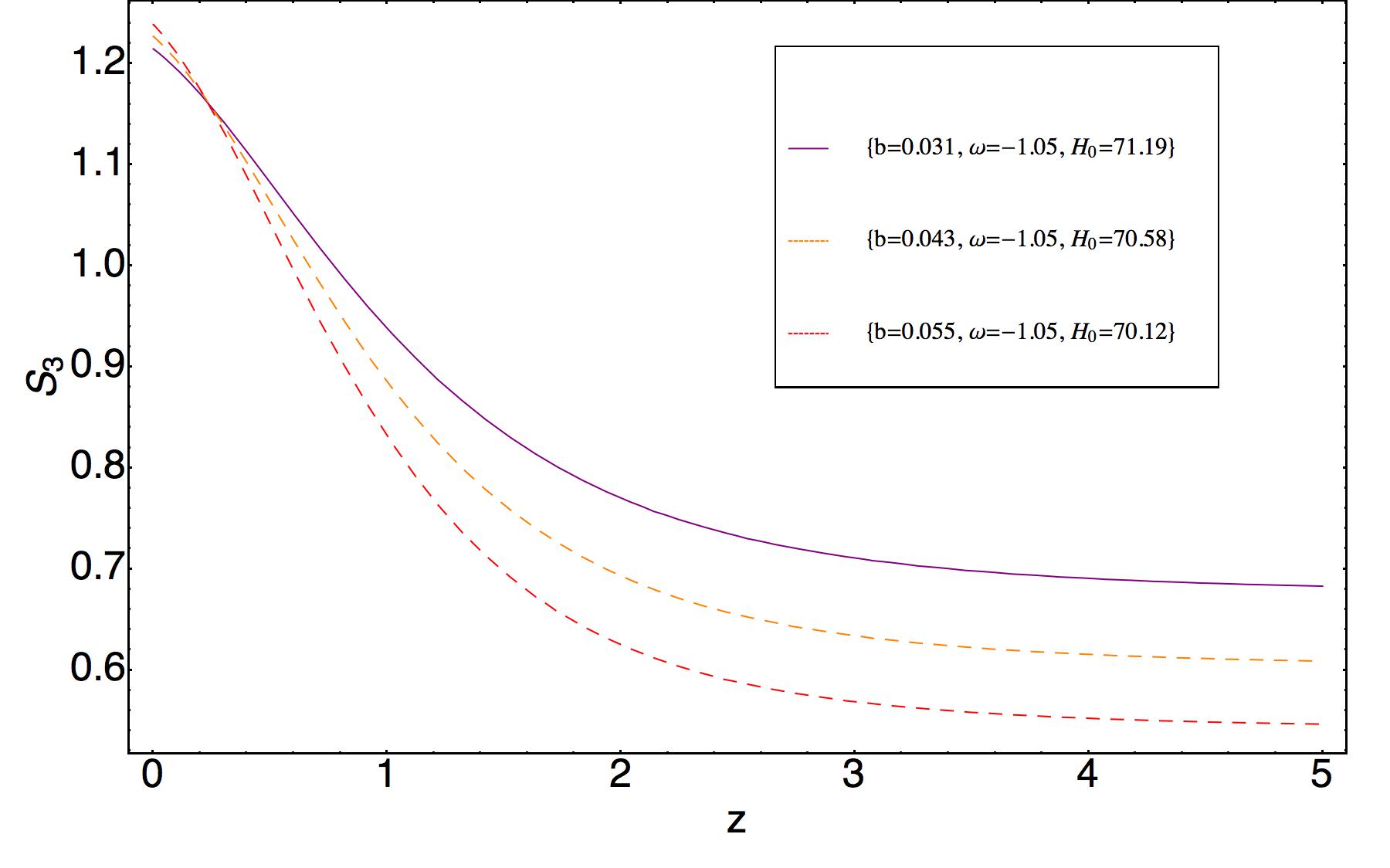}  \\
 \end{array}$
 \end{center}
\caption{The top panel represents graphical behavior of the deceleration parameter $q$, $\Omega_{de}$~(solid lines) and $\Omega_{dm}$~(dashed lines). The bottom panel represents graphical behavior of $Om$ and $S_{3}$ parameters. The case corresponds to the model of the universe with two component fluid, when the non-gravitational interaction is given by Eq.~(\ref{eq:Q4})}
 \label{fig:Fig5}
\end{figure}

\begin{figure}[h!]
 \begin{center}$
 \begin{array}{cccc}
\includegraphics[width=80 mm]{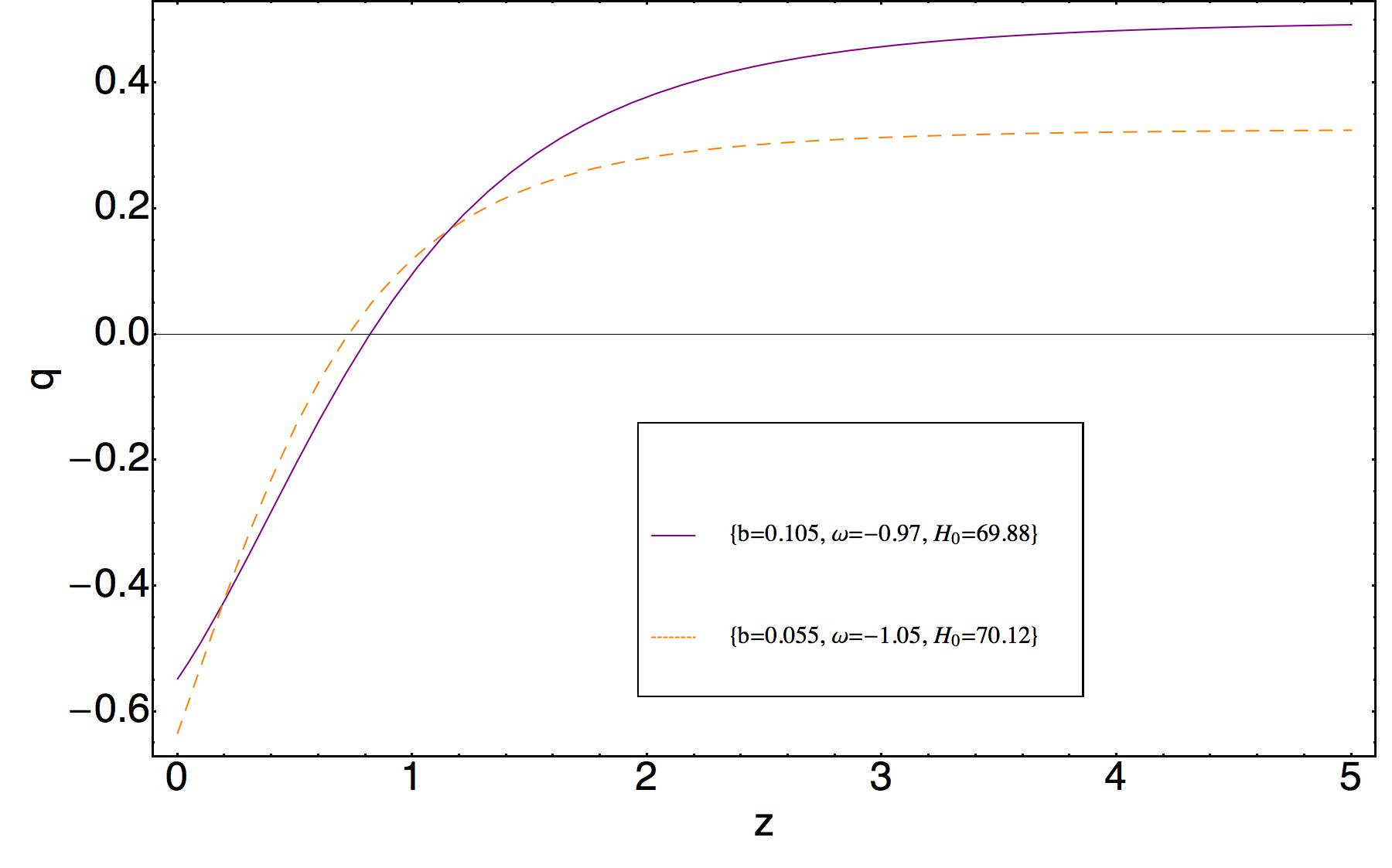}  &
\includegraphics[width=80 mm]{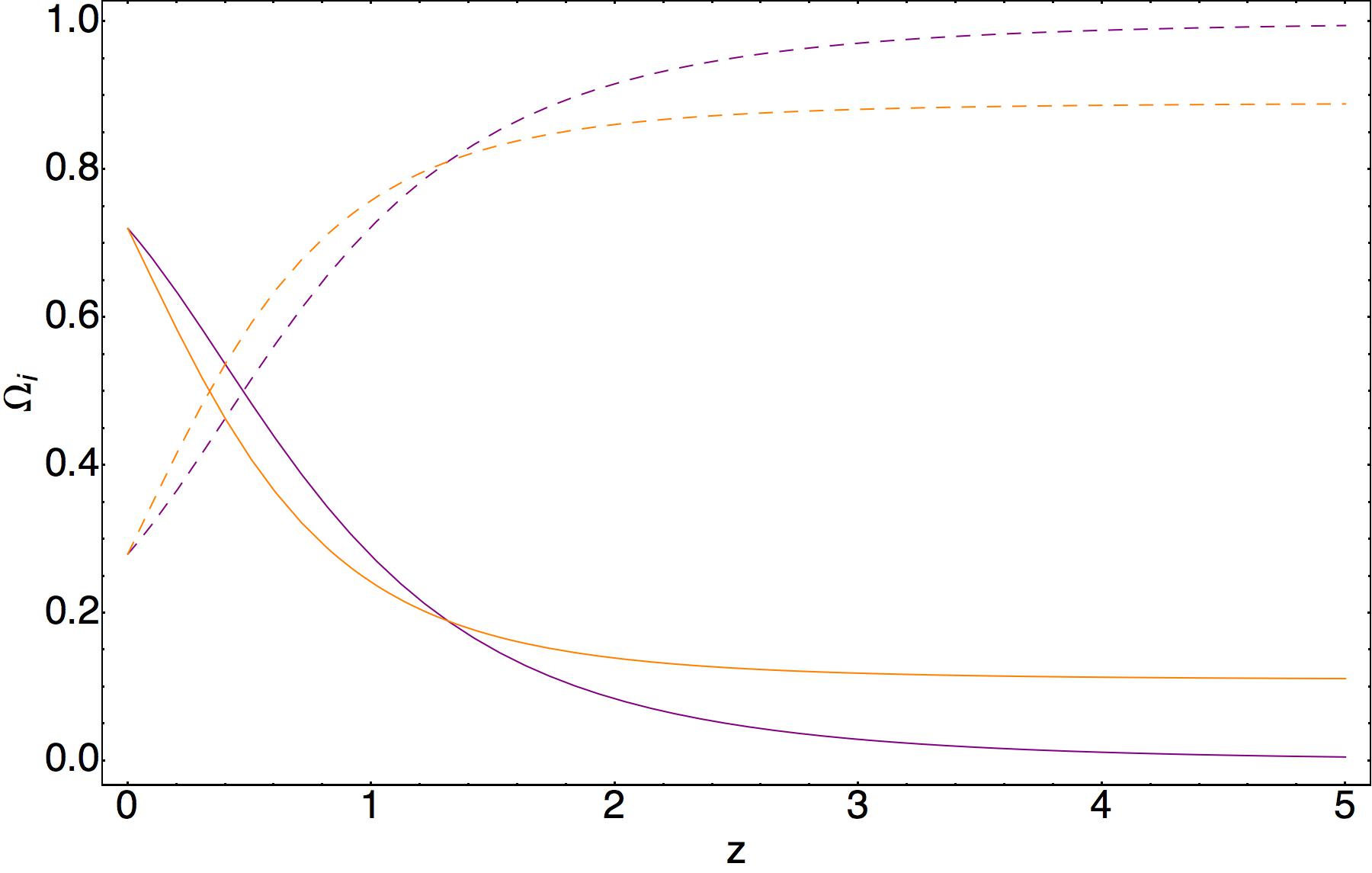}  \\
\includegraphics[width=80 mm]{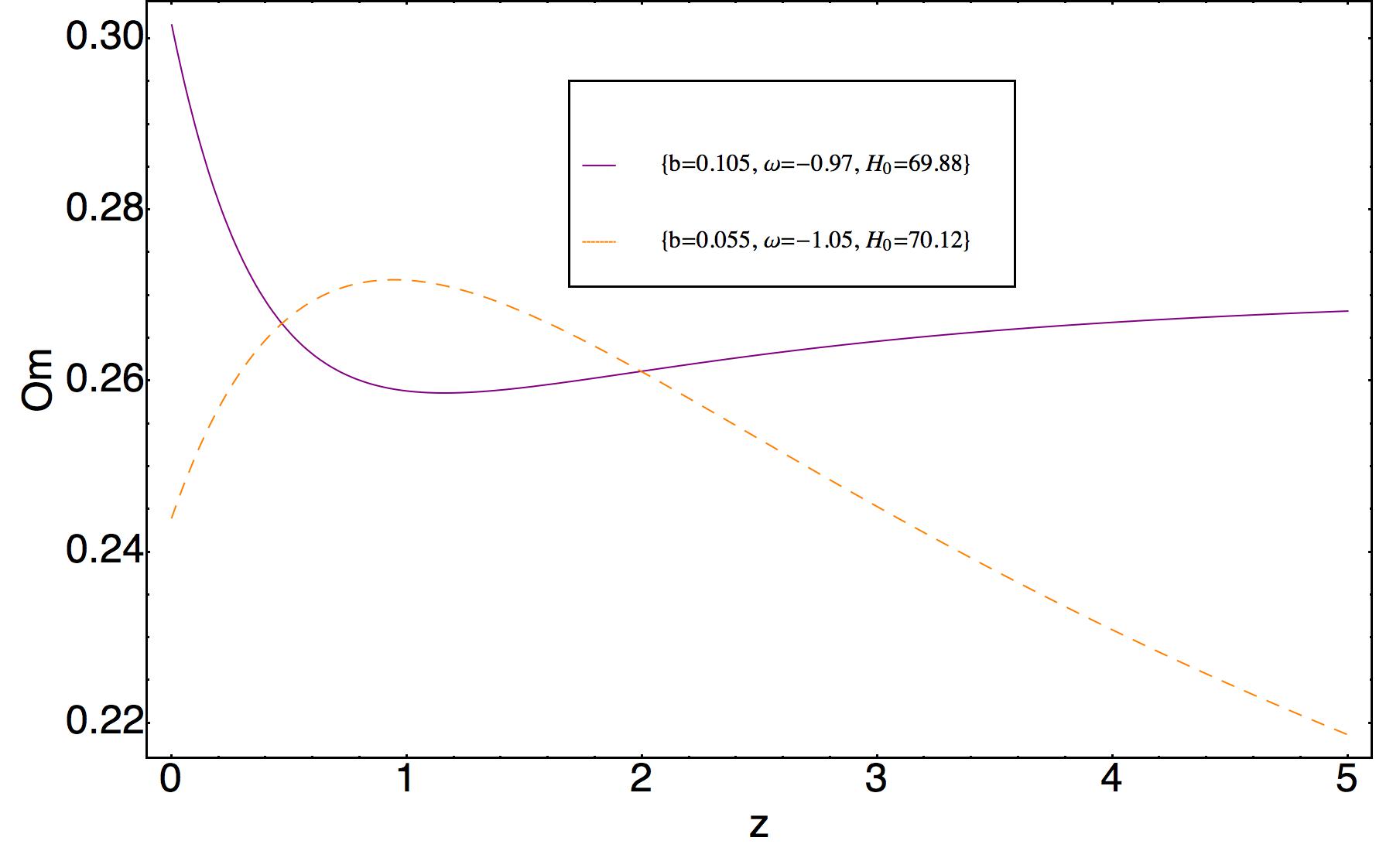}  &
\includegraphics[width=80 mm]{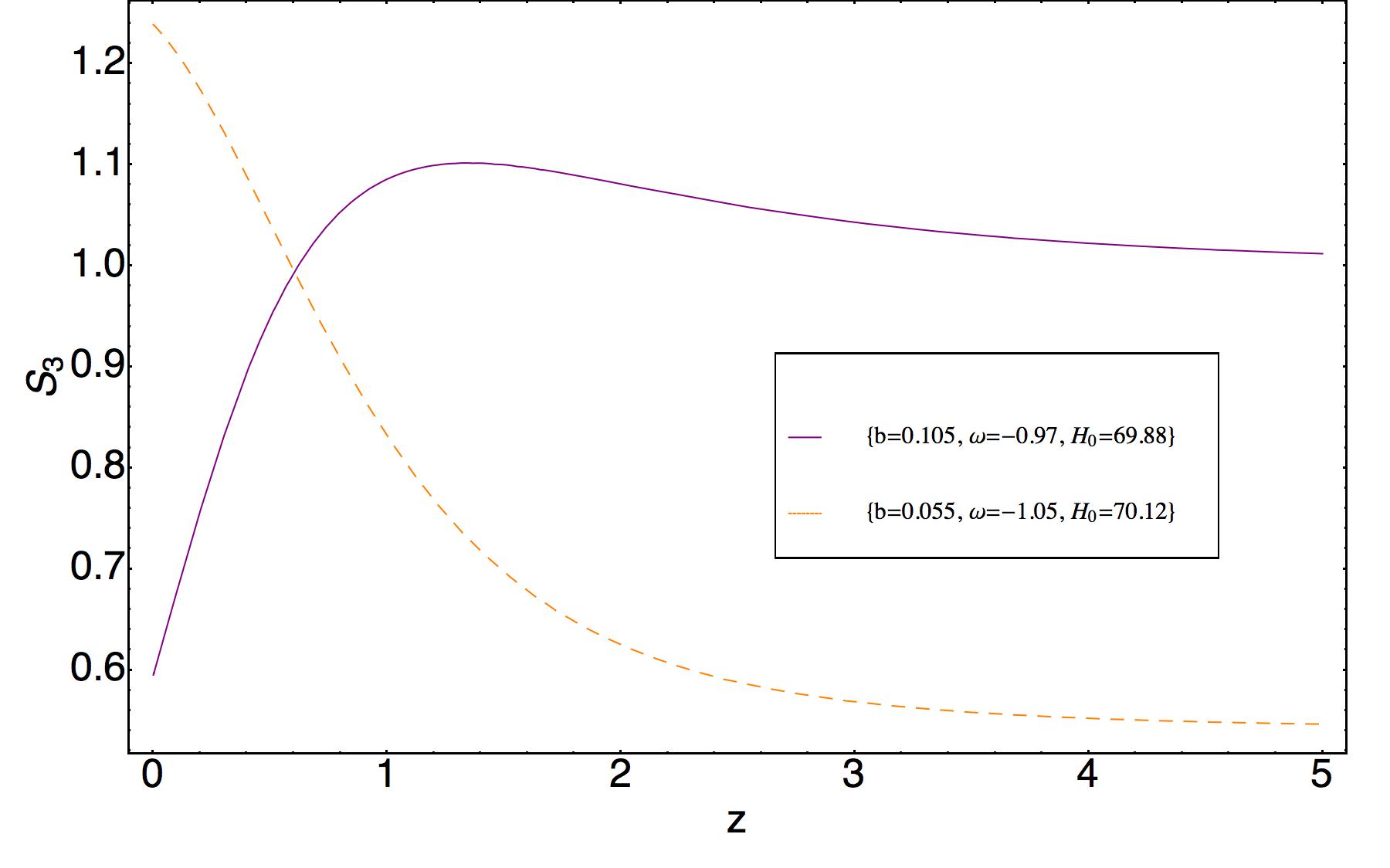}  \\
 \end{array}$
 \end{center}
\caption{Graphical behavior of the deceleration parameter $q$, $\Omega_{de}$~(solid lines), $\Omega_{dm}$~(dashed lines), $Om$ and $S_{3}$ parameters for the two models of the universe, when the non-gravitational interaction is given by Eq.~(\ref{eq:Q1})~(blue line) and Eq.~(\ref{eq:Q4}), respectively}
 \label{fig:Fig6}
\end{figure}

In this section we present our study on the cosmological model, where the non-gravitational interaction between dark energy and dark matter is given by the following expression:
\begin{equation}\label{eq:Q3}
Q = 3 H b \rho_{dm} \log \left[ \frac{\rho_{de}}{\rho_{dm}}\right].
\end{equation}
Moreover, such interaction in our case leads to energy transition from dark energy to dark matter, while the interaction
\begin{equation}\label{eq:Q4}
Q = 3 H b \rho_{dm} \log \left[ \frac{\rho_{dm}}{\rho_{de}}\right],
\end{equation}
will indicate energy transition from dark matter to dark energy. In this case, the analysis reveals an interesting fact. Using $\chi^{2} = \chi^{2}_{OHD} + \chi^{2}_{BAO}+ \chi^{2}_{\mu} + \chi^{2}_{SGL}$ statistical technique, it turns out that more favorable is to have transition of the energy from dark matter to dark energy. In particular, the scanning of the parameters space shows, that for the model with $\Omega^{(0)}_{dm} = 0.26$ when $H_{0} = 71.19$, $\omega = -1.05$ and $b = 0.03$ we obtain the best fit, characterized by $\chi^{2} = 783.7$. On the other hand, for $\Omega^{(0)}_{dm} = 0.27$ the best fit is observed for $H_{0} = 70.58$, $\omega = -1.05$ and $b = 0.043$~($\chi^{2} = 780.0$), while for the model with $\Omega^{(0)}_{dm} = 0.28$, it is observed when $H_{0} = 70.12$, $\omega = -1.05$ and $b = 0.055$~($\chi^{2} = 778.6$). Graphical behaviors of the deceleration parameter $q$, $\Omega_{de}$, $\Omega_{dm}$ are presented in Fig.~(\ref{fig:Fig5}). Graphical behavior of $Om$ and $S_{3}$ parameters are presented in the bottom panel of Fig.~(\ref{fig:Fig5}). To finalize this section, in Fig.~(\ref{fig:Fig6}) we present the graphical behaviors of $q$, $\Omega_{de}$, $\Omega_{dm}$, $Om$ and $S_{3}$ parameters for the models presented in subsection~\ref{ss:11} and here, characterized by the smallest $\chi^{2}$ for each case. For instance, from comparison of the behavior of the deceleration parameters we see, that for the model described by the interaction term given in Eq.~(\ref{eq:Q4}) it is a constant for higher redshifts. On the other hand, for the model with the interaction Eq.~(\ref{eq:Q1}) the transition redshift is smaller and the present day value of the deceleration parameter is higher compared to the model with the interaction given in Eq.~(\ref{eq:Q4}). From the right plot of the top panel of Fig.~(\ref{fig:Fig6}) we see a consequence of this on the redshift dependent behavior of $\Omega_{de}$ and $\Omega_{dm}$. Comparison of $Om$ and $S_{3}$ parameters is presented on the bottom of Fig,~(\ref{fig:Fig6}). It is evident, that both parameters are applicable to distinguish the differences of both models under comparison.

\section{Discussion}\label{sec:Dis}

It is well known, that existence of non-gravitational interaction can make theoretical models work better. Nevertheless, there is no a fundamental theory explaining the origin of the connection of non-gravitational nature between two dark components. Dark components operate on different scales, which makes the problem more complicated. Various new phenomenological parameterizations of the interaction are considered recently. There is an increasing interest towards non-linear and non-linear sign changeable interactions to improve the background dynamics described by the general relativity. Two types of new parameterizations of non-gravitationtal interaction between cold dark matter and dark energy described by the barotropic fluid equation of state are suggested here. To constrain the models $\chi^{2}$ statistical technique is applied. When the form of the interaction is given by Eq.~(\ref{eq:Q1}), the model with the energy transfer from dark energy to dark matter is preferred from the observational data (see section~\ref{sec:M1}). If the energy transfer from dark matter to dark energy is allowed, then in such universe, dark energy should be a phantom. Moreover, the value of the equation of state parameter describing dark fluid is bellow the value obtained by PLANCK 2015 experiment. 

Study of the models with non-gravitational interaction, i.e. interaction is given by Eq.~(\ref{eq:Q3}), shows that the energy transfer from dark matter to dark energy is preferred by the observational data (see section~(\ref{sec:M2})). The constraint on the equation of state parameter describing dark energy in this case providing the best fit is in a good agreement with the result obtained by PLANCK 2015 experiment. In this model the phantom line crossing is possible, while in the model in section~\ref{sec:M1} only quintessence universe is occurred.

The differences between the models considered in sections~\ref{sec:M1} and~\ref{sec:M2} are studied by the minimal $\chi^{2}$~(Fig.~\ref{fig:Fig6}). In both cases graphical behavior of the $Om$ and $S_{3}$ parameter~(statefinder hierarchy analysis) are analyzed. Both tools reveal all differences between the considered models. For instance, $Om$ analysis shows, that for higher redshifts the cosmological model considered in section~\ref{sec:M1} becomes $\Lambda$CDM standard model, while for lower redshifts there is a huge difference between the considered and $\Lambda$CDM models. The $Om$ analysis indicates two redshifts, where the properties of the models considered in sections~\ref{sec:M1} and ~\ref{sec:M2} are the same, while $S_{3}$ parameter indicates just only one redshift. This features indicates existence of differences between these two analyses. This is another subject of research, which will be reported elsewhere. The considered parametrizations of the interaction found to be supported by the observational data of a certain type, on the other hand in considered cosmological models the cosmological problems are solved. Moreover, it can be checked that obtained best fit values for the parameters provide results satisfying $Omh^{2}$ analysis as well.

\end{document}